\documentclass[preprint,12pt]{elsarticle}
\usepackage{amssymb}
\usepackage{amsmath}
\usepackage{adjustbox}
\usepackage{longtable}
\usepackage{enumitem}

\journal{Human Factors in Health Factor} 

\usepackage{tabularx}
\usepackage{booktabs}
\usepackage{graphicx}
\usepackage{hyperref}
\usepackage{hhline}
\usepackage{svg}
\usepackage{geometry}

\usepackage{tabularray}

\usepackage{subcaption}
\usepackage{enumitem}

\usepackage{float} 
\usepackage{soul,color}
\usepackage[utf8]{inputenc}
\usepackage{chngcntr}
\usepackage{xcolor}
\definecolor{editCol}{rgb}{0.0, 0.0, 255.0}

\usepackage{pbox}
\usepackage{multirow}
\usepackage{array}
\usepackage{makecell}
\usepackage{afterpage} 
\usepackage[title]{appendix}
\usepackage{subcaption,graphicx}
\usepackage{ragged2e}
\usepackage{url}
\usepackage{hyperref}

\usepackage{booktabs, caption, makecell}

\usepackage{threeparttable}

\newcolumntype{L}[1]{>{\raggedright\let\newline\\\arraybackslash\hspace{0pt}}m{#1}}
\newcolumntype{C}[1]{>{\centering\let\newline\\\arraybackslash\hspace{0pt}}m{#1}}
\newcolumntype{R}[1]{>{\raggedleft\let\newline\\\arraybackslash\hspace{0pt}}m{#1}}

\begin{document}
\begin{frontmatter}

\title{Focus on the Experts: Co-designing an Augmented Reality Eye-Gaze Tracking System with Surgical Trainees to Improve Endoscopic Instruction}


\author{Jumanh Atoum}
\author{Jinkyung Park}
\author{Mamtaj Akter}
\author{Nicholas Kavoussi}
\author{Pamela Wisniewski}
\author{Jie Ying Wu}

\begin{abstract}
The current apprenticeship model for surgical training requires a high level of supervision, which does not scale well to meet the growing need for more surgeons. Many endoscopic procedures are directly taught in the operating room (OR) while the attending surgeon and trainee operate on patients. The need to prioritize patient care limits the trainees' opportunities to experiment and receive feedback on their performance. Augmented reality (AR) has the potential to increase efficiency in endoscopic surgical training, but additional research is critical to understanding the needs of surgical trainees to inform the design of AR training systems. Therefore, we worked with 18 surgical trainees to understand the strengths, limitations, and unmet needs of their current training environment and to co-design an AR eye-gaze tracking system based on their preferences. Trainees emphasized the need to practice the 2D to 3D mapping needed to properly familiarize oneself with the anatomy of patients to prepare for real surgery. The trainees felt that an AR-based eye gaze tracking system would be a useful supplemental training method that would improve their learning in OR cases without detracting from patient care. To tailor the AR system to their needs, they co-designed features to improve their ability to track the attending surgeon's eye gaze and to provide a real-time, interactive system. Our results are valuable in shaping the endoscopic training modules by generating user-informed guidelines to design future collaborative AR-based eye-gaze tracking systems.
\end{abstract}




\begin{keyword}
Augmented Reality \sep Eye-gaze Tracking \sep Co-design Focus Group \sep  Qualitative Thematic Analysis
\end{keyword}

\end{frontmatter}

\section{Introduction}
The U.S. healthcare industry faces a projected surgeon shortage of 15,800 to 30,200 by 2034. In response, the government has introduced legislation to increase the number of medical students, leading to a growing demand for surgical training. Unfortunately, the need to train these students places additional strain on surgeons~\cite{parsons2011}. Most surgical training takes place in a clinical setting through an apprenticeship model, where trainees first observe experts performing procedures and then perform procedures under their direct supervision. The apprenticeship model, while facing challenges due to the surge in medical students and trainees, remains irreplaceable~\cite{lee2018can} as it fosters collaboration and interaction, both inside and outside the operating room (OR)~\cite{lee2018can}. 

Maintaining the collaboration in the apprenticeship model is necessary, and emerging technologies such as virtual reality (VR)/ augmented reality (AR) can be leveraged to enhance the collaboration between surgical trainees and experts. These technologies have been integrated into surgical training programs to teach both behavioral and non-behavioral skills, including technical, cognitive, and domain-specific abilities~\cite{o2024say}, in the OR~\cite{jing2021eye}. We focus on AR-based assistive tools as they can be seamlessly integrated into the OR, unlike VR tools~\cite{barsom2016systematic}. We propose to study design choices in integrating AR into surgical training in the OR through AR Head-Mounted Displays (AR-HMDs) and support collaborative interactions~\cite{hamacher2016application,semsar2019trainees, fussell2004gestures}. In particular, we propose to use eye-gaze as an hands-free way for the surgeon to provide additional visual guidance to the trainee. Acar et al. demonstrated through preliminary studies that AR can improve endoscopic kidney stone identification in a simulated OR environment~\cite{acar2023Intraoperatie}. Endoscopic treatment of kidney stones is particularly challenging to learn, as all training currently occurs in the OR. Although the minimally invasive approach offers significant patient benefits, such as reduced recovery time and blood loss, it also demands intensive and specialized surgical training~\cite{zakeri2020, fransson2022training}. Inadequate training has been linked to suboptimal outcomes, with 23\% of patients requiring reoperation within 20 months due to residual stone fragments~\cite{brain2021natural}, underscoring the critical need for more effective training strategies~\cite{kimchi2020minimally}. To enhance the AR design to improve training efficiency in the OR, we pose the following research questions (RQs):

\begin{itemize}
    \item \textbf{RQ1:} \textit{What are the strengths, weaknesses, and unmet user needs associated with the current endoscopic surgical training settings used by surgical trainees?}
    \item \textbf{RQ2:} \textit{How would surgical trainees design an AR-based training system to overlay experts' eye gaze for endoscopic training?}
\end{itemize}

To answer our RQs, we conducted six co-design sessions with 18 surgical trainees. Through qualitative analysis of responses, we make these contributions:  
\begin{itemize} 
\item Determined the limitations of the current surgical training environment and where AR can improve training efficiency
\item Demonstrated key design considerations and new feature co-designs for AR-based visual guidance in surgical training from the perspectives of trainees
\item Developed design considerations for AR visual features to improve the endoscopic training process between surgical experts and trainees within the OR
\end{itemize}

Our findings can inform the development of future AR-based assistive systems to facilitate collaboration between expert surgeons and trainees. Therefore, our paper makes advances in human factors in the healthcare community.

\section{Background}
\subsection{VR-/AR-based Systems in Surgical Training} 
VR is a cost-effective tool for surgical training, offering greater interaction with virtual environment than traditional, screen-based simulators and improving skill acquisition~\cite{mao2021immersive, dhar2023scoping}. VR technology fully immerses users in a computer-generated environment, blocking out the real world. VR enables safe practice before real-life cases, enhancing patient safety~\cite{mazur2022novel}. Studies confirm VR's effectiveness in teaching transferable skills~\cite{seymour2008vr, walbron2020virtual, lohre2020effectiveness}. However, VR-based training modules can only be used for training outside of the OR since users have no access to what happens outside the virtual world~\cite{suresh2023role}. Some VR applications let multiple users collaborate in real time with headsets to manage a virtual patient in environments like a simulated hospital room, addressing both medical management and teamwork~\cite{abelson2015virtual}. However, these applications lack realism~\cite{gallagher1999virtual}, which creates uncertainty about how easily users can apply the skills they gain to the OR. Researchers have yet to conduct enough studies to evaluate how well these skills transfer to real surgical settings.

AR is emerging in surgery for preoperative planning and execution across specialties like pediatrics, orthopedics, oncology, and neurosurgery~\cite{gehrsitz2021cinematic, lu2022applications, scherl2021augmented, qi2021holographic}. Unlike VR, AR overlays digital data such as text and images on physical objects in the real world, integrating seamlessly into the OR~\cite{barsom2016systematic}, allowing for real-time interaction while maintaining sterility in the OR. For instance, Qi et al. showed that AR-neuronavigation achieves 81.1\% lesion localization accuracy, comparable to standard methods, while not increasing the overall operating time compared to standard navigation~\cite{qi2021holographic}. Furthermore, Dennler et al. reported that surgeons show high satisfaction on a 100-Likert scale when using AR in the OR~\cite{dennler2021augmented}. AR also aids in visualizing internal anatomy for emergency care~\cite{castelan2021augmented}. While these studies highlight the increasing demand for AR applications, their primary focus lies in hologram manipulation and planning processes. Prior works on AR-based intraoperative applications have also focused on providing a form of visual guidance with little to no focus on training between surgical experts and trainees in the OR~\cite{eom2022, tong2024development}. While AR tools aid training, they can cause eyestrain, dizziness, and fatigue if not user-tailored~\cite{iskander2019using}.

\subsection{AR-based System for Eye-Gaze Tracking in Endoscopic Training}

Eye-gaze metrics are widely used to assess surgical skills and distinguish experts from novices~\cite{law2004eye, richstone2010eye, khan2012analysis, li2023eye, nespolo2023platform}. Vine et al showed that surgical trainees can improve their performance and learning experience by actively adopting expert gaze strategies in a collaborative setting~\cite{vine2012cheating}. Li et al. showed that surgical trainees and experts exhibit different eye-gaze patterns~\cite{li2023eye}. Further introducing expert eye gaze for guidance improves overall performance~\cite{acar2023Intraoperatie}. 

Given the potential drawbacks of AR-based systems, evidence-based guidelines to improve AR-visual guidance, such as eye-gaze pointers, in endoscopic surgical training are under-studied. To the extent of our knowledge, this is the first study conducted for designing a user-centered AR-based guidance tool for endoscopic kidney stone surgery training in the OR. We aim to provide design considerations for AR guidance to maximize the efficiency of surgical training.

\section{Methods}

\subsection{Study Overview}
The primary purpose of our work is to conduct a formative evaluation of an AR-based eye-gaze tracking prototype to generate evidence-based guidelines. This can inform user-centered and effective design of future AR-based endoscopic surgical training systems. 

We combined semi-structured interviews with a co-design session to allow surgical trainees to provide in-depth insights into their training needs and visual representations of how they would redesign the eye-gaze tracking feature for AR-based surgical training systems. This approach effectively addressed our research questions by merging a retrospective interview approach, which elicits users' thoughts and experiences, with a generative methodology (i.e., co-design), where researchers collaborate with users to envision features that meet the users' needs~\cite{muller1993, spinuzzi_methodology_2005}. Co-design approaches vary depending on the design stage, ranging from contextual inquiry (ideation) to collaborative prototyping (design) to iterative redesign of existing systems~\cite{muller1993}. Our study falls between the ideation and design stages, allowing us to explore participants' understanding of the problem space and their design ideas, extending beyond the authors' preconceptions.

\subsection{Participant Recruitment and Demographics}
Our study was approved by the university's Institutional Review Board (IRB). Participants were required to be currently training or interning at our university's hospital and could participate individually or in groups. We included participants from both the urology and otolaryngology departments, comprising residents who had completed all their departmental rotations and medical students who had finished 2-3 rotations. This diversity led to varying levels of hands-on OR experience with kidney stone endoscopy, though all participants except from the otolaryngology department had observed the procedure. While all participants were from the same hospital, residents completed rotations at different institutions, and some had prior internship experience at other hospitals, resulting in diverse exposure to surgical training modules. None had experience with AR training.
We advertised the study through recruitment emails and word-of-mouth within the university's urology and otolaryngology departments. Recruitment took place from March 2023 to June 2024, with 18 surgical trainees ultimately recruited. Participant demographics are shown in ~\autoref{tab:demographics}. Among the participants, eleven identified as female and seven as male, with their post-graduation years (PGY) ranging from zero to six years of practice.

\begin{table}[ht]
\begin{threeparttable}[t]
    \centering
     \footnotesize
    \caption{Demographics of the participants. 0 years indicate not graduated. T$\#$ indicates Trainee, M$\#$ indicates Medical Student, and PGY indicates post-graduation years of practice}
    \label{tab:demographics}
    \begin{tabular}{ m{2cm}  m{3.5cm}  m{2.5cm}  m{5.4cm} }
        \hline \\
         \textbf{Group ID} &  \textbf{Participants ID
        }& \textbf{Gender} & \textbf{PGY$\#$} \\[0.3cm] \hline 
         G1 & T1, T2, T3 & F, F, M & 4 yrs, 4 yrs, 4 yrs \\
         G2 & T4 & F & 3 yrs \\
         G3 & T5, T6  & M, M & 4 yrs, 4 yrs \\
         G4 & T7, T8, T9 & F, M, F & 2 yrs, 3 yrs, 6 yrs \\
         G5 & T10, T11, T12 & M, F, M & 6 yrs, 3 yrs, 1 yrs \\
         G6 & M1, M2, M3, M4 & F, F, F, M& 0 yrs, 0 yrs, 0 yrs, 0 yrs \\
        G7 &M5 &F &0 yrs \\
        G8 &M6 &F &0 yrs \\
        \hline
    \end{tabular}
\end{threeparttable}
\end{table}

\subsection{Study Procedure}
As shown in~\autoref{tab:demographics}, we conducted a total of eight sessions, ranging between one to four participants per session. Each study session consisted of two distinct phases: 1) A semi-structured interview and 2) A co-design session. Each group session took approximately 30 minutes to an hour for participants to complete. There was no monetary compensation provided to participants. We describe the two distinct research phases in more detail below.

\subsubsection{Semi-structured interviews} 

To structure the interview questions, we first formulated our high-level research questions among the research team members, who have diverse backgrounds in computer science, surgical robotics, human-computer interaction co-designs, and surgery. The questions about the surgical training environment were guided by our surgical collaborator, the fourth author, while the questions related to HoloLens (our choice of HMD technology) were inspired by the questions in the existing literature~\cite{evans2017evaluating, baashar2023towards}. \autoref{table:InterviewQuestionspartA} presents sample questions.
\begin{table*}[!ht]
 \centering
 \footnotesize
\caption{Structure of the Co-design Sessions with Sample Questions and Tasks} 
  \label{table:InterviewQuestionspartA}
\begin{tabular}{ |p{3cm}|p{11cm}|  }
 \hline
\textbf{Research Phase} & \textbf{Sample Questions} \\ \hline

\textbf{Phase 1: \newline Semi-structured interview} & 
\setlist{nolistsep}
\begin{itemize}[leftmargin=0.3cm, noitemsep]
    \item \textit{Can you explain in detail all training modules you have come across so far?}
    \item \textit{How do you feel about your current training setup?}
    \item \textit{What are things about your current training setup you like?}
    \item \textit{What aspects of your current training setup you want to change?}
\end{itemize} 
 \\
\hline
\textbf{Phase 2: \newline Co-design session} & 
\textbf{Presented pictures of HoloLens}
\setlist{nolistsep}
\begin{itemize}[leftmargin=0.3cm, noitemsep]
    \item  \textit{Would you want to use the HoloLens during procedures or in training? If not, why not?}
    \item \textit{What kind of support do you think you would need for using the HoloLens for this purpose?}
    \item \textit{Are there any concerns you might have regarding the use of HoloLens in your training?}
\end{itemize}
\textbf{Showed video of overlayed eye-gaze data during simulated surgery}
\setlist{nolistsep}
\begin{itemize}[leftmargin=0.3cm, noitemsep]
    \item  \textit{Write down what you observed from the video you just saw.}
    \item \textit{Based on what you saw, what do you think of overlaying eye-gaze data on the recorded videos?}
    \item \textit{Was there something that stood out based on the snapshots given to you? If so, please tell me about what stood out to you.}
\end{itemize}  

\\
\hline

\textbf{Phase 2: \newline Co-design session continued} &

\textbf{Task 1:} Use the whiteboard, markers, and papers to sketch what you think is the most valuable information in the snapshot.
\setlist{nolistsep}
\begin{itemize}[leftmargin=0.3cm, noitemsep]
    \item \textit{What do you think would be the best way to deliver direction and hand movement information using any approach other than communicating it verbally?}
\end{itemize} 
\textbf{Task 2:} Sketch the approach you came up with from the previous question. Swap doodles and sketches with the group.
\setlist{nolistsep}
\begin{itemize}[leftmargin=0.3cm, noitemsep]
    \item  \textit{What is the first thing you noticed when you saw the [other person's] sketch? (This can be anything).}
    \item \textit{Discuss as a group what each sketch was pointing to and what you thought about this approach.}   
     \item \textit{Was it easy to identify what the sketch was pointing to? Why (not)?}
    \end{itemize} 
\textbf{Task 3:} In the light of non-verbal guidance, write down some positive aspects and areas of improvement for this sketch.  
\\
\hline

\end{tabular}
\end{table*}
We conducted a pilot study - separate from the eight main group studies - with an expert surgeon (a co-author) to simulate the structure and flow of the planned user studies. The objective of this pilot was to develop a deeper understanding of the surgical training environment and refine our research approach. The expert surgeon, who has prior experience with AR surgical training, collaborated with us to evaluate and revise our interview questions. This process helped us identify relevant applications and limitations of AR in the context of endoscopic kidney stone procedures. The pilot study provided valuable insights into areas where surgical trainees typically encounter challenges and guided our exploration of how AR could offer meaningful support in these contexts. By combining the clinical expertise of the surgeon with the AR design knowledge of our computer science research team, we ensured that the subsequent user studies were both contextually grounded and technically informed. 

\subsubsection{Co-design sessions with Design Probe} 

\begin{figure*}
\begin{subfigure}[ht]{0.3\linewidth}\centering
  \includegraphics[width=\textwidth, height=1.55in]{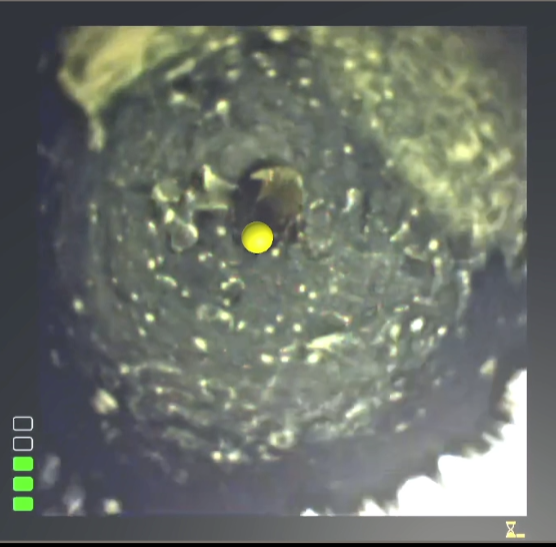}
\caption{Screenshot of video}
\end{subfigure}
\hspace{1em}
\begin{subfigure}[ht]{0.5\linewidth}\centering
  \includegraphics[width=\textwidth, height=1.55in]{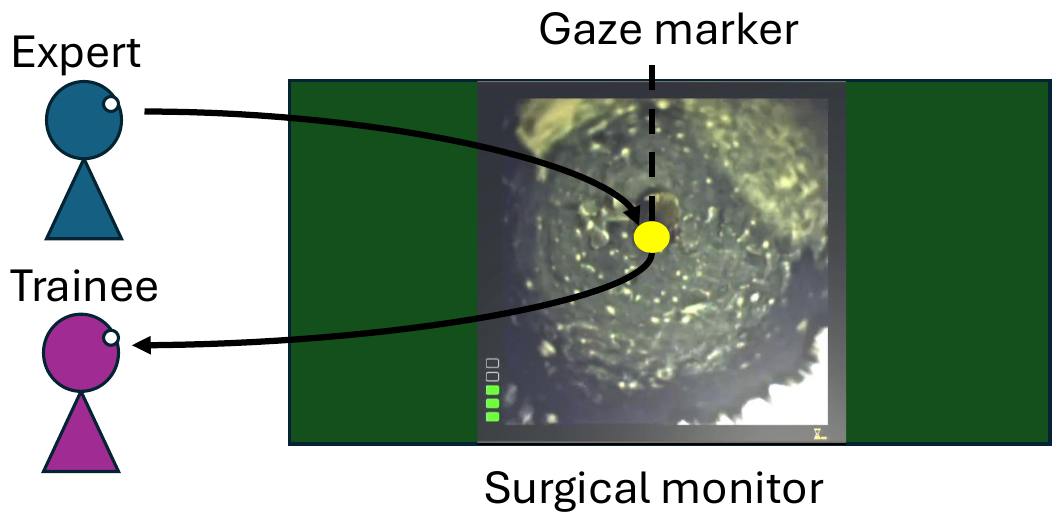}
\caption{Expert-trainee communication via visual guidance} 
\end{subfigure}
\caption{\textbf{(a)} A screenshot of the video displayed to the participants, overlaying a yellow sphere on a pre-recorded video of a simulated stone identification task. \textbf{(b)} Illustration of how expert and surgical trainees would communicate through visual guidance in an AR-based eye gaze tracking.} 
\label{fig:system}
\end{figure*}

\textbf{\textit{Design Probe:}} To conduct co-design sessions with surgical trainees, we used a pre-recorded video of a simulated stone identification task, featuring an overlaying pointer (\autoref{fig:system}), as a design probe. This allowed participants to envision using an AR-based eye-gaze tracking system for surgical training. This AR eye-gaze tracking system will dynamically follow the experts' eye-gaze and be projected on a 2D screen to provide visual guidance for the surgical trainees. Design probes enable participants to interact with system prototypes in a tangible context, encouraging open-ended feedback and providing insights into their thought processes as they complete tasks~\cite{wallace2013, Shneiderman2016}.

The two-minute video served as a non-interactive design probe and a demonstrative tool. To create this demonstration, we recorded eye-gaze data from an expert surgeon wearing a HoloLens while performing kidney-stone identification on a kidney phantom. The expert’s gaze was visualized as a yellow sphere superimposed on the endoscopic video, illustrating a potential method of real-time visual guidance. This design probe allowed trainees to visualize how following an expert's eye gaze might look and consider the features needed for an AR-based visual guidance system. Our co-design study aimed to explore how AR systems could address the specific needs of surgical trainees. A screenshot from the video demo is shown in~\autoref{fig:system}.

\textbf{\textit{Co-design Tasks:}} First, we showed surgical trainees pictures of the Microsoft HoloLens 2 (Microsoft Corporation, Redmond, WA) to assess their familiarity and impressions of the AR equipment. For example, we asked participants about their understanding of the HoloLens features, its potential use cases, and how those features might be applied in both surgical procedures and training contexts. These initial discussions helped gauge their baseline knowledge of AR-based systems and informed the direction of our follow-up questions. Next, we asked how they would use AR in training and what kind of support they would want. After presenting the video, we asked participants if the design probe was relevant and useful for their everyday training. We followed up with questions on different techniques for providing visual guidance in the OR using AR-based systems and similar technologies.

Finally, we asked participants to imagine themselves in training, with the attending surgeon’s eye gaze cast on the surgical monitor, guiding them through the kidney stone identification procedure, as shown in \autoref{fig:system} (a). While imagining this scenario, we asked them to brainstorm visual assistive features focused on casting the surgeon’s eye gaze. Participants were provided with a whiteboard and colored markers to sketch their designs. Throughout the co-design session, we encouraged participants to think aloud and discuss their ideas within the group\cite{guan_validity_2006}. After completing their sketches, participants walked us through their designs, explaining the rationale behind each feature and annotating their sketches.

\subsection{Data Analysis Approach}
The first author conducted and transcribed all co-design sessions and took photos of the design artifacts drawn on the whiteboard. A grounded thematic analysis was used to identify themes for RQ1. The first author reviewed the data, created initial codes, and met with co-authors to refine the codes. New codes were added as needed, and after the coding process, the first author and the second-to-last author grouped the codes into themes. The themes for RQ1, along with their codes and illustrative quotes, are presented in \autoref{table:codebook1}. For RQ2, both the transcripts and photos of the design artifacts were coded. \autoref{table:codebook2} shows the grouped features identified in the co-design sessions. Given the open-ended nature of the questions, some answers were double-coded, resulting in a code count higher than the number of participants. For instance, the ``strengths'' theme had a code count of 57 due to participants providing multiple perspectives on the varying forms of verbal and visual guidance received during training.

\begin{figure*} [htb]
\begin{subfigure}[ht]{.26\linewidth}\centering
  \includegraphics[width=\textwidth, height = 1.2in]{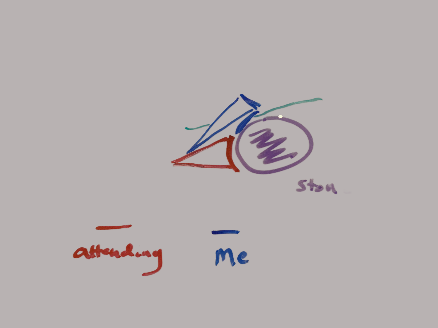}
  \caption{ Gaze tracking}
\end{subfigure}
\hspace{1em}
\begin{subfigure}[ht]{.26\linewidth}\centering
  \includegraphics[width=\textwidth, height = 1.2in]{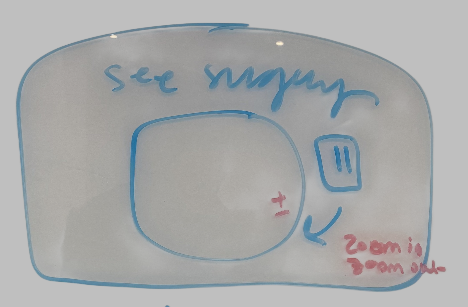}
  \caption{ Mixed learning approaches}
\end{subfigure}
\hspace{1em}
\begin{subfigure}[ht]{.38\linewidth}\centering
  \includegraphics[width=\textwidth, height = 1.2in]{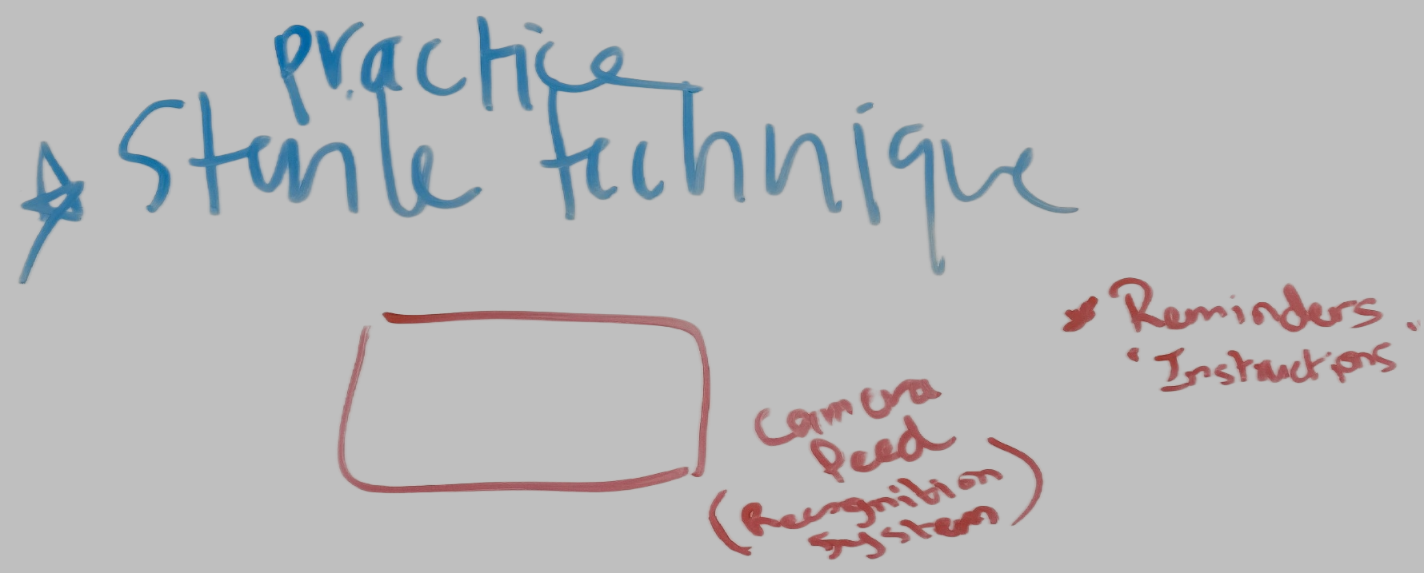}
  \caption{ Directed real-time feedback}
\end{subfigure}
\caption{ Co-designs: (a) Visual guidance provided through experts' markers annotations in a dual robotic training system. (b) Mixed learning to allow surgical trainees to retrieve pre-recorded cases at any time. (c) Remotely accessing the HMD-embedded camera for real-time feedback.}
\label{fig:theme_1_3codes}
\end{figure*}

\section{Results}
\subsection{Strengths, Weaknesses, and Unmet User Needs for Surgical Training Environment (RQ1)}
Our surgical trainees had experience with various training environments, including open surgery training, robotic console training simulations, and dual robotic consoles for training under experts' supervision. The robotic console provides a VR simulator for robotic surgery, while the dual console allows trainees to observe experts during robotic surgery on patients. The robotic consoles provided training exclusively for robotic surgeries so they cannot be used for training endoscopic skills, but we list them because they provided the trainees VR and AR training exposure and informed their co-designs. However, their exposure to these systems varied based on their post-graduation years \autoref{tab:demographics}. 
\begin{table}[!ht]
\centering
 \footnotesize
\caption{Codebook for Strengths, Weaknesses and Needs of Current Surgical Training Environment (RQ1) }
\begin{tabular}{|>{\raggedright}p{3.8cm}|>{\raggedright}p{3.3cm}|@{ }p{6.5cm}|}
\hline
\textbf{Themes} & \textbf{Codes} & \textbf{Quotes} \\
\hline
\multirow{3}{4cm}{\textbf{Strengths: Surgical trainees saw varying training approaches in different surgical training modules.}} & Visual Guidance (n*= 17) & \textit{``We usually can see where the surgeons want us to go...''}-T5 \\ \cline{2-3} 
& Skill Assessment \newline (n = 14) &  \textit{``After each task, the console shows a report of how we have done in the training practice and keeps track of our progress.''}-T7 \\ \cline{2-3} 
& Directed Real-Time Feedback (n = 15) &  \textit{``Attending surgeon can take over the console and direct us..''}-T1 \\  \cline{2-3} 
&  Mixed Learning Opportunities \newline (n = 11) &  \textit{``We are able to practice outside of OR at our own convenience..''}-T3 \newline \\  \cline{2-3}
\hline 
\multirow{2}{4cm}{\textbf{Weaknesses: Surgical Trainees identified disconnects between surgical training environments and OR procedures.}}& Lack of Endoscopic Training Outside OR (n = 15) &   \textit{“I feel like it would be nice to practice driving ureteroscope around a kidney...... before doing it in OR
case.''}-T10\\  \cline{2-3} 
& Absent Surgical Realism (n = 12) &  \textit{``It is hard to conceptualize how the organ looks like and how the organ will deform by just observing how phantoms behave in the simulators.}''-M1 \\ \cline{2-3} 
& Overly Simplistic Training (n = 11) &  \textit{``I feel like most of the training we get from the robotic console is too general, for example, tying a knot…''}-T4 \newline \\
\cline{2-3}
\hline
\multirow{2}{4cm}{\textbf{Unmet Needs: Surgical Trainees expressed the need for integrating virtual 3D anatomies for collaborative learning.}} & 2D to 3D Mapping (n = 15) &   \textit{``I guess we can have a map of the kidney at the bottom right of the monitor that shows a pointer of where we are in real-time.''}-T2 \\  \cline{2-3} 
& Anatomy Knowledge \newline (n = 14) &  \textit{``It would be helpful to show the anticipated anatomy vs imaging vs what they are seeing.''}-T2\\ \cline{2-3} 
& Pre-operative Practice (n = 12) &  \textit{``I wish I could have played a bit with the robotic console right before doing a case''}-T6 \newline \\ \hline
\multicolumn{3}{l}{*n represents the number of participants contributed to codes} \\
\end{tabular}
\label{table:codebook1}
\end{table}
\subsubsection{Varying assistive approaches in different surgical training modules} 
Almost all of our participants shared that they benefited from the multimodal learning existing in the training environments. They learned from visual guidance in OR, with attending surgeons directing them to target surgical locations, highlighting specific anatomical information. Almost all of our participants (n = 17, 94\%) felt that pointers for \textbf{visual guidance} are an important component in both robotic console and endoscopic surgery training environments. An example is the co-design in ~\autoref{fig:theme_1_3codes} (a) where one participant sketched what the visual guidance looks like in a dual robotic training system. Pointers show the difference between tools navigated by a surgical trainee versus an expert.
\begin{quote}
\textit{``Dual robotic consoles provide a better ergonomic environment, the attending does not need to get up and point at the screen they can move a small pointer from their side. It is really helpful when they need to show us where to go next.''} - T1
\end{quote}
The majority of the participants (n = 16, 89\%) brought up the importance of \textbf{skill assessment report} provided by the 2D-surgical and robotic console simulations. The skill report provided a detailed analysis of their performance within the selected training module, such as the time taken to perform each task and the efficiency of their movement, ranging from comprehensive assessments of complete procedures to focused evaluations of specific skill sets. The compiled information provides an indicator of a surgical trainee's progress toward meeting the standards set by the attending surgeon or training program. Participants noted that automated assessment tools helped them track skill development and boosted their confidence by offering progress reports.
\begin{quote}
    \textit{``After each task, the console shows a report of my performance. It displays score of performance skills, including overall time, the economy of the overall movement. It also keeps a record of previous performances, so you can tell if you are doing well.''} - T7
\end{quote}
Most of our participants (n = 15, 83\%) also saw strengths in \textbf{real-time and interactive feedback} they received from the attending surgeons within the OR. Surgical trainees often shared that OR training environments facilitated supervised training, which was exclusive to the OR training. Participants emphasized that this training approach boosted their confidence during OR procedures and played a key role in refining their surgical techniques. Medical students (n = 6, 33\%) saw the importance of having supervision for assessing their performance when practicing sterile techniques~\autoref{fig:theme_1_3codes} (c). The sketch shows a red box representing the camera field-of-view as seen by the medical student; the camera feed is transmitted over a video conferencing software system, allowing remote supervision.
\begin{quote}
  \textit{``We usually get direct feedback when we are stuck on tasks, surgeon can easily catch our mistakes, take over the console, and direct us.''} - T1
\end{quote}
Participants highlighted the individualized learning environments and peer-to-peer training opportunities. More than half of participants (n = 11, 61\%) expressed appreciation for the \textbf{mixed learning opportunities} provided in all outside-the-OR training environments, including open, robotic console, and endoscopic simulations. They underscored the effectiveness of peer-to-peer, allowing them to practice skills without extending procedure times. The participants cited self-learning in ~\autoref{fig:theme_1_3codes} (b) through watching pre-recorded cases. This can be used in group discussions between peers while providing multiple features such as pausing video and zooming in and out.
\begin{quote}
    \textit{``Learning is limited in OR since patients are prioritized over teaching, having simulators to use in our own time is good for practicing.''} - T3
\end{quote}
Overall, all participants mentioned that the built-in features of different training modalities provided them with accurate surgical and anatomical information, both within and beyond the OR.

\subsubsection{Disconnects between surgical training environments and OR procedures}
While our surgical trainees generally held a favorable opinion regarding the existing training systems, they perceived several gaps. Almost all of the participants (n = 15, 83\%) voiced their concerns about the \textbf{lack of proper endoscopic training outside of the OR}. Participants shared that they felt hesitant when performing their first endoscopic OR case. This is because they lacked prior practice to develop the needed skills to acquire proficiency with instrument navigation.
\begin{quote}
     \textit{``Usually the first time we perform endoscopies is in OR, we always struggle at first and so I feel like it would be nice to practice driving [endoscope] around a kidney model before experiencing it in OR case.'' - T12 }
\end{quote}
Our participants also frequently discussed the use of model anatomy for training, called phantoms, for practicing simple knot-tying, suturing, and needle-passing outside the OR. The majority of our participants (n = 12, 66\%) expressed that \textbf{phantoms lacked realism} as they did not mimic real organs. Therefore, they found it challenging to gain effective anatomical and surgical insights solely by practicing on phantoms. This difficulty led to a lack of confidence in applying the practiced skills in real OR cases.
\begin{quote}
    \textit{``It is hard to conceptualize how the organ looks like and how the organ will deform by just observing how phantoms behave in the simulators. Also, it is almost always that we build the conceptualization in the OR after we perform on few OR procedures.''} - M1
\end{quote}
Around two-thirds of our participants (n = 11, 61\%) noted that the skill sets taught by the robotic consoles and surgical simulators were \textbf{overly simplistic} and did not translate well into the real OR cases they encountered. The skill sets were broad and rudimentary and mastering the skills did not contribute significantly to their long-term surgical proficiency. This was evident when the participants reflected on the long duration it took them to become comfortable and confident performing procedures in the OR under supervision.
\begin{quote}
     \textit{``the skill sets acquired from some of the tasks we practice on aren't easily transferred, for example when we practice knot-tying, the thread used is thick and not used in the procedure of course. Unless we go into OR and do it ourselves, there is nothing that mimics it accurately''} - M3
\end{quote}
\begin{figure*} 
\begin{subfigure}[ht]{.35\linewidth}\centering
  \includegraphics[width=\textwidth, height = 1.2in]{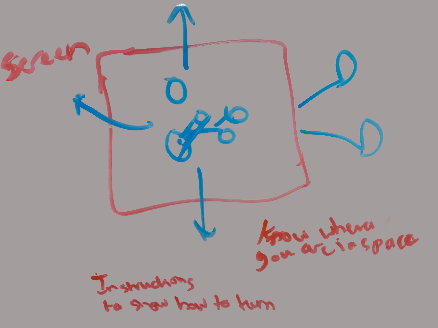}
  \caption{ 2D to 3D mapping}
\end{subfigure}
\hspace{1em}
\begin{subfigure}[ht]{.35\linewidth}\centering
  \includegraphics[width=\textwidth, height = 1.2in]{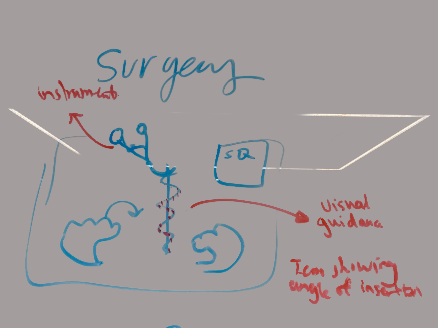}
  \caption{Pre-operative practice}
\end{subfigure}
\hspace{1em}
\begin{subfigure}[ht]{.21\linewidth}\centering
  \includegraphics[width=\textwidth, height = 1.2in]{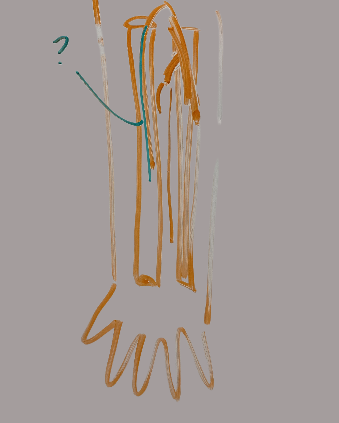}
  \caption{Anatomy knowledge}
\end{subfigure}
\caption{ Unmet needs are met through AR: (a) Mapping the orientation of the anatomy in 2D images to its orientation in the world during OR cases. Surgical screen is a red box, the blue drawings show the internal organ, and the arrows represent the orientation of the user. (b) Preoperative practice on suturing can be provided through projecting step-by-step suturing instructions. (c) Familiarizing surgical trainees with anatomies by projecting holograms of hidden anatomies.}
\label{fig:theme_3_3codes}
\end{figure*}

\subsubsection{Need for integrating virtual 3D anatomies for assistive learning}
All of our participants felt a need to build a conceptualization of organs and organ interaction before going into OR. For example, most participants (n = 15, 83\%) emphasized the importance of \textbf{projecting 2D images as 3D holograms} of the anatomy for a deeper grasp of anatomy and its changes over time, which aids decision-making during unexpected events in the OR. Surgical trainees noted that the 2D-surgical and robotic console simulations failed to effectively teach anatomical structure identification, lacking realism and not bridging the gap between theoretical knowledge and real-world OR scenarios. To address this issue, participants expressed the need for a 3D map mechanism during endoscopic procedures. This would help them develop navigation skills and correlate 2D images on the surgical monitor with the endoscope's position and orientation, ultimately enabling them to construct a 3D map of the organ in their minds. An example sketched during the co-design shown in ~\autoref{fig:theme_3_3codes} (a) describes a need for real-time position and orientation of the tool-tip with respect to patients' bodies. This directly ties to the difficulty of building 3D maps in their heads.
\begin{quote}
    \textit{``In some cases, where we need to insert our tool into the kidney, we need to know exactly where to hit, this usually requires a lot of 3D thinking, to know the right location and not miss it.''} - T2
\end{quote}
In addition to projecting 3D holograms, three-fourths of our participants (n = 14, 77\%) wanted \textbf{anatomical holograms}, as tissues can rapidly take on an unfamiliar appearance after dissection under different surgical scenarios. They often suggested integrating various dissected anatomical models in the endoscopic training simulators to better resemble real tissue deformation and behavior in the human body and provide experts with different examples that they can teach from. \autoref{fig:theme_3_3codes} (c) shows an example for how co-designs can support learning anatomy through a hologram/ physical model of a hand. The drawings inside represent the different arteries, tendons, and veins that trainees must learn.
\begin{quote}
    \textit{``Anatomy under different circumstances looks unfamiliar to us compared to experts leading to a lot of anonymity in terms of what to focus on For example, if we are dissecting an organ outside of human body it will behave differently than inside, due to pressure surrounding organs.''} - M4
\end{quote}
Furthermore, our participants (n = 12, 67\%) felt they needed more \textbf{observational learning}. All of these surgical trainees expressed that warm-up sessions structured for the specific procedures would help them feel more confident and less stressed compared to the current training approach. They often said that even pre-recorded videos that describe common operation procedures would help. This would remind them of what to expect, what tasks to do, and how to do them efficiently. This is an example of the need for system and user collaboration rather than relying solely on expert and surgical trainee interaction. \autoref{fig:theme_3_3codes} (b) describes an example video for a suturing procedure. The blue box contains a sketch of two hands, scissors, and a line with a zigzag showing suture imprints. In the sketch, the feedback is given in the smaller box as a set of instructions as envisioned by the participants. 
\begin{quote}
    \textit{``I think having raw surgical videos to watch before going into the OR is helpful. It will be even more helpful to walk through the video with the attending or another resident after the surgery to ask more questions that we usually don’t get the chance to during the procedure.}'' - T6
\end{quote}
In general, surgical trainees were more familiar with robotic and laparoscopic simulators but faced challenges with intraoperative endoscopic training, highlighting the need to replicate real-world OR settings in training.

\subsection{Design Considerations for AR-based Endoscopic Training Applications (RQ2)}
In the co-design phase, all 18 surgical trainees expressed interest in AR-based training systems but emphasized that these systems would complement, not replace, their current training tools. Some trainees noted that AR could address gaps in existing setups, as T1 mentioned:
\begin{quote}
\textit{``I think it would be good to have those AR features to complement the existing training setup since it can improve in some areas that the current training setup lacks.''} - T1
\end{quote}
About one-third of participants (n = 5, 28\%) expressed concerns about needing additional training to use AR system features. Another (n = 5, 28\%) was concerned about the practicality of using the bulky HoloLens for training.
\begin{quote}
\textit{``I am worried that the HoloLens would be too heavy or cause fatigue after wearing for some time, similar to the VR displays''} - T10
\end{quote}
In their co-designs, surgical trainees primarily focused on two key features: 1) eye-gaze tracking and 2) real-time interactive holograms, as detailed in \autoref{table:codebook2}, to enhance guidance during practice cases. Below, we describe these features and their specific properties designed by the trainees.
\begin{table}[!ht]
\centering 
 \footnotesize
\caption{Feature Analysis for AR-based Endoscopic Training Applications (RQ2)}
\label{table:codebook2}
\begin{tabular}{|p{2.8cm}|p{3.8cm}|p{6.8cm}|}
\hline
\textbf{Feature} & \textbf{Properties} & \textbf{Description} \\
\hline
\multirow{2}{3cm}{\textbf{Tracking Attending Surgeons' Eye-gaze}} & Simple geometric shape (n* = 8) & Marker should not have more than one edge \\ \cline{2-3}
& Small marker size (n = 8) & The pointer should be small enough so as not to cover any important anatomical details \\ \cline{2-3}
& Contrasting color (n = 8) & Use colors that contrast with the inside of the human body. \\ \cline{2-3}
& Visibility (n = 5) & Hide/unhide the marker when necessary\\ \hline
\multirow{3}{3cm}{\textbf{Real-time learning tools for experts and surgical trainees}} & Real-time localization \newline  (n = 6) & Show a map where the trainee is inside the organ \\ \cline{2-3}
& Interactive anatomical holograms (n = 7) & Draw and erase on the anatomical features, and interact with anatomical holograms\\ \cline{2-3}
& Show trails (n = 4) & Leave a trail to mark areas that have been visited by the trainee inside the organ\\ \hline
\multicolumn{3}{l}{ 
*n represents the number of participant groups contributed to codes} \\
\end{tabular}  
\end{table}

\begin{figure*} 
\begin{subfigure}[ht]{.45\linewidth}\centering
  \includegraphics[width=\textwidth, height = 1.45in]{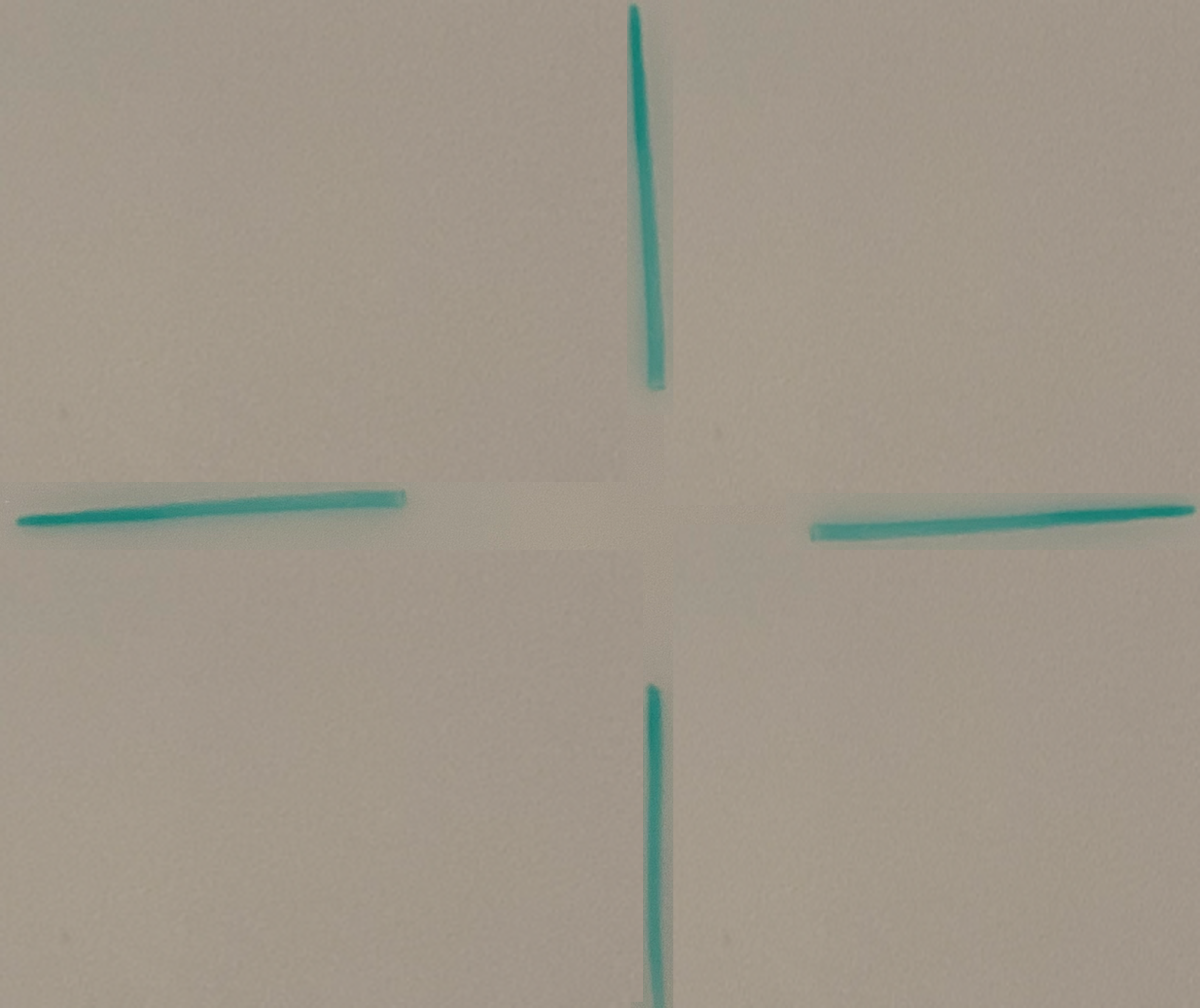}
  \caption{ Simple geometric shape}
\end{subfigure}
\hspace{1em}
\begin{subfigure}[ht]{.5\linewidth}\centering
  \includegraphics[width=\textwidth, height = 1.45in]{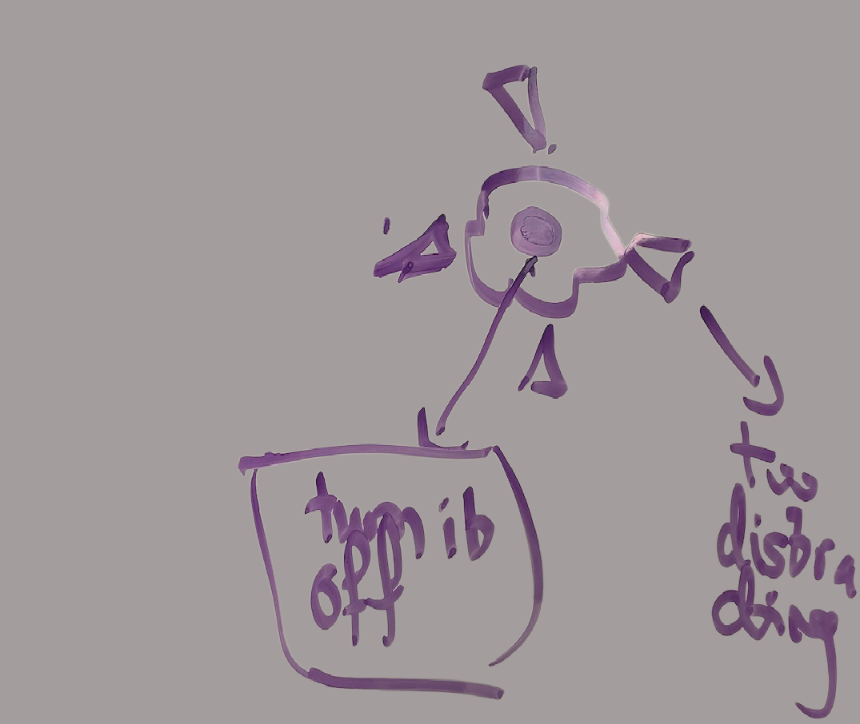}
  \caption{ Transparency control}
\end{subfigure}

\caption{Desired properties of eye-gaze tracking pointers: (a) Simple geometric shape: target-like shape to force focus on the elements in the center of shape whilst not being too distracting, and (b) Visibility control: set the pointers' transparency and hide/unhide the marker.}
\label{fig:feature-2}
\end{figure*}
\subsubsection{AR features to track attending surgeons' eye-gaze} 

All of our participant groups suggested incorporating an AR feature to track the attending surgeon's eye gaze during the OR procedures. Since co-designs are the result of group efforts, we analyze the frequency of AR features appearing based on the number of participating groups, not individual participants. The AR feature would allow trainees to understand the expert's visual focus and aid navigation through different anatomies. Trainees often cited the inefficiency of verbal communication, where multiple tasks are discussed at once, and the difficulty of manipulating the endoscope. These challenges were worsened by the limited realism of training outside the OR, making real OR cases the only opportunities for learning and feedback. 
\begin{quote}
    \textit{``It is important to know where they are pointing or where they want us to go, do we go up into the right hole...''} - T4
\end{quote}
To support eye-gaze tracking in AR-based training systems, participants designed several features. They suggested using simple geometric shapes with few or no edges for the pointer representing the attending surgeon's eye gaze. All groups (n = 8) emphasized that static, non-blinking shapes would be less distracting during training.  Others suggested game-like target pointers with the area of interest at the center. Ultimately, all groups settled on simple geometric shapes for the pointer. An example of this design is shown in ~\autoref{fig:feature-2} (a), where the center of the cursor indicates the point of interest. 
\begin{quote}
     \textit{`` think a simple target-like cursor, similar to a gaming cursor, should be sufficient.''} -  T1
\end{quote}
All eight surgical trainee groups suggested making the pointer small to avoid obstructing anatomical features or surgical information. However, they emphasized that the pointer size should be relative to the organs and the surgical camera's field of view. Trainees noted that an appropriately sized eye-gaze marker would reduce cognitive load, preventing the need to search for the marker on the monitor. 
\begin{quote}
    \textit{``Pointer should be small enough to not cover any important anatomical details.''} - T8
\end{quote}
In addition to marker size, all eight participant groups also brought up the importance of choosing the right \textbf{color of the eye-gaze pointer}. The colors they proposed were blue and green to contrast with the color of the tissues within organs and suggested avoiding colors such as yellow or red. 
Trainees also often suggested incorporating features to adjust the marker color based on the part of the body they are performing on. For example, choosing a contrasting color to the bone tissue would be different from that of a soft organ tissue. 
\begin{quote}
    \textit{``I think pointers have to have a color that contrasts to the inside of human body, else you cannot differentiate between pointer and organ.''} - T10
\end{quote}
Most participant groups (n = 5, 62.5\%) designed features to hide or adjust the transparency of the marker during training. They emphasized that a persistent marker on the screen could distract them from focusing on critical areas of tissue or instruments and may block anatomical features. To address this, they suggested the ability to remove or control the transparency of the marker, as demonstrated by group G4 in ~\autoref{fig:feature-2} (b).
\begin{quote}
    \textit{``Dynamic eye tracking can be really distracting, having a continuous pointer on the screen can become an obstacle, so giving us control over hiding and unhiding would be better...''} - T6
\end{quote}
\begin{figure*} 
\begin{subfigure}[ht]{.43\linewidth}\centering
  \includegraphics[width=\textwidth, height = 1.2in]{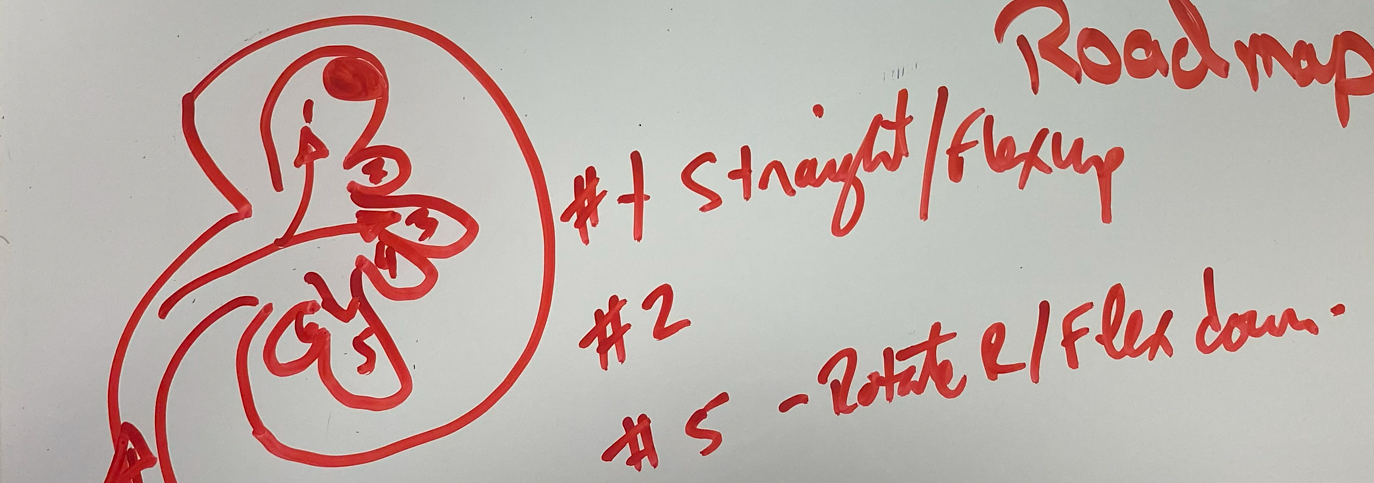}
  \caption{ Holographic 2D map of the kidney}
\end{subfigure}%
\hspace{1em}
\begin{subfigure}[ht]{.24\linewidth}\centering
  \includegraphics[width=\columnwidth, height = 1.2in]{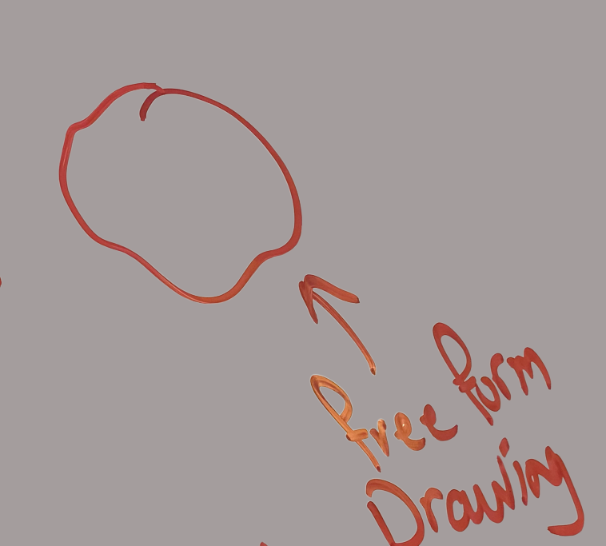}
  \caption{ Interactive hologram}
\end{subfigure}
\hspace{1em}
\begin{subfigure}[ht]{.24\linewidth}\centering
  \includegraphics[width=\textwidth, height = 1.2in]{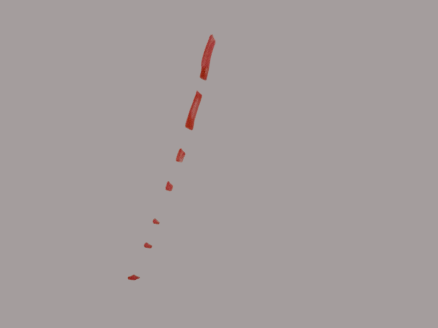}
  \caption{Trail guidance}
\end{subfigure}
\caption{Features for AR Guidance Resources: (a)  Holographic 2D map of the kidney shows a map where the trainee is inside the organ in real-time., (b)  An example of interactive holographic feature providing free-form drawing on the surgical monitor and interacting with anatomical holograms and (c)  Trail shape left by experts to guide navigation}
\label{fig:feature-1}
\end{figure*}

\subsubsection{Surgical trainees designed AR features that can facilitate real-time learning tool via visual guidance}
Participants discussed features enabling real-time, two-way AR guidance between experts and trainees. They designed real-time visual AR support features to improve training efficiency and overcome the limited exposure to OR cases.
\begin{quote}
    \textit{``I think a real-time system should be able to take our inputs and tell us where we are or would tell us how we should move the endoscope based on our movements...''} - T3
\end{quote}
All participant groups (n = 8) designed organ maps for real-time localization, showing the current endoscope tip position inside the anatomy. This feature would help trainees develop mapping skills, visualize maneuvers, and understand the surgical path. Participants suggested the map could be a detailed anatomical figure, a computed tomography image, or a simple diagram illustrating possible paths to the target area. ~\autoref{fig:feature-1} (a) shows an example of a kidney map designed by group G3.
\begin{quote}
    \textit{``..visualize the scope in real time in actual space rather than in 2D space on the screen, to help us know where we are at in the kidney and decide on our next trajectory.''} - T11
\end{quote}
The majority of participant groups (n = 7, 87.5\%) designed interactive anatomical holograms, enabling trainees to draw on the surgical monitor using erasable holograms. This feature allows collaboration between experts and trainees, where both can highlight anatomical details and correct errors. Additionally, the trainees often brought up how overlaying images on real anatomical structures would allow them to better visualize how the organ would look after deformation. They emphasized that holograms should not obstruct the view of the organs. An example, designed by Group G4, is shown in ~\autoref{fig:feature-1} (b).
\begin{quote}
    \textit{``I would like to be able to draw and erase on the hologram while others can see it, also, I guess the holograms should be transparent to not hide any anatomical features.''} - T3
\end{quote}
Half of the participant groups (n = 4) designed ``pathways'' or \textbf{trails} in the areas of the organ that had been explored. Participants emphasized the value of visualizing the attending surgeon's path during a procedure, particularly when collaborating for the first time to navigate the organ. 
~\autoref{fig:feature-1} (c) shows a drawing by Group 2 for how the trail should look.
\begin{quote}
    \textit{``Leaving a trail to mark areas that have been visited and showing the trajectory made by the attending can help us remember the trajectory better, especially when we can see it.''} - T5
\end{quote}
Overall, the surgical trainees designed features providing better real-time guidance and visualization resources that would assist them during training.

\section{Discussion} 

\subsection{Optimizing Endoscopic Training by Addressing Unmet Needs}
Our results highlight the importance of visual guidance for surgical training and integrating more visual guidance into endoscopic training in OR. Participants perceived that the current training systems provide limited, summative skill assessment reports that do not show where their performance degraded or improved. Further, learners at the stage of unconscious incompetence will not know that they need help with, pushing for the need for real-time monitoring and experts' feedback. This goes in line with the participants' conceptualization of the need for real-time skill assessment. Participants emphasized the importance of receiving real-time, targeted feedback and expressed the need for a navigation tool that provides directional guidance to help surgical trainees—especially during endoscopic training—accurately reach their target location. The need for systems that provide real-time and targeted feedback stems from the limited availability of expert supervision in many training environments~\cite{nieboer2019recruiting}. This highlights the importance of developing collaborative training tools that foster interaction between experts and surgical trainees, as shown in prior work~\cite{brennan2008coordinating,lee2017improving}, especially in domains like endoscopic training, where such guidance is often insufficient~\cite{acar2023Intraoperatie}.

Our findings reinforce and expand upon prior work~\cite{7} that identified the lack of surgical realism in current training environments. Participants highlighted the difficulty in transferring learned skills from practicing outside to performing in surgical rooms. The limitation in surgical realism highlights the need to design AR training modules that incorporate environmental factors to simulate the OR environment and provide training simulations for high-risk procedures. Such modules need 3D models of organs for surgical education. Further research is needed to design 3D anatomical models with better texturization to improve surgical realism.

\subsection{Leveraging AR-based Eye Gaze Tracking System for Surgical Training}

Our study revealed that current intraoperative endoscopic training is limited by restricted practice opportunities outside the OR and the need to minimize procedure time. Participants emphasized the value of seeing attending surgeons' eye-gaze patterns for visual guidance in endoscopic surgeries, highlighting the shortcomings of verbal communication in the OR. Verbal instructions can be overwhelming due to the complexity of information conveyed~\cite{rossnagel2004lost}. Tracking the attending surgeon's eye gaze and projecting it as a hologram marker on the HMD was identified as key to improving communication. Our findings outline design considerations for using HMDs and AR applications to address communication limitations and enhance collaboration and training in the OR. The dependence of surgical trainees on verbal communication in OR training exemplifies existing collaboration between them and experts~\cite{garosi2020concerns} and the need for additional visual guidance tools to improve training efficiency.

As surgical trainees are often taught by different experts, variations in instructional styles and inherent biases can influence how procedures are performed. This highlights the need for more objective methods of skill assessment. We propose using eye gaze tracking to supplement traditional skill assessments. By analyzing pre-recorded eye gaze, we can generate objective skill assessments, such as a report detailing task load and focus levels at different procedure stages. Previous work has shown varying stress levels during different parts of procedures, making eye gaze a valuable metric for measuring skill ~\cite{li2023eye}. 

\subsection{Improve Endoscopic Training through Visual Guidance}

One key finding from our study is the potential for real-time visual guidance during localization and navigation in endoscopic procedures. Surgical trainees rely on expert-provided markers for visual guidance during robotic surgery, which is unavailable in endoscopic procedures. Participants expressed a preference for visual guidance through a virtual pointer/marker to track where the expert was looking. This aligns with prior work emphasizing markers for creating a collaborative environment~\cite{feng2019communication}, but our study goes further by empirically demonstrating the need for eye gaze as a visual guidance tool.

Participants also suggested annotating static holograms and surgical monitors during endoscopic procedures as another form of visual guidance. Previous research showed that expert surgeons do not prefer multi-user input for annotations during peer discussions~\cite{maria2023supporting, johnson2011exploring}. Therefore, we propose deploying visual annotators as a teaching tool for explaining procedures and revisiting certain anatomies rather than relying on expert annotations. 

Trainees raise the concern that if the AR system is not properly designed, it can obstruct the trainees' field of view. A key design consideration is knowing when to provide visual guidance and who should control it. Prior work has generally given control to experts without considering the preferences of surgical trainees~\cite{feng2019communication}. We advocate for a more trainee-centered training system, where the surgical trainee has control over the visibility of the guidance tool, including the ability to hide or unhide visual elements. Future work could focus on creating a more flexible ``control'' property, where both the expert and trainee can control visual elements, promoting a more discussion-based training approach.

\subsection{Guidelines for Designing AR-based Features for Eye Gaze Tracking and Supporting Real-time Guidance for Endoscopic Training}

Based on our findings, we propose the following design considerations for creating real-time and interactive visual guidance tools to improve endoscopic surgical training. Further research is needed to validate their utility experimentally.

\begin{itemize}
     \item Improve training within OR by \textbf{tracking and projecting the attending surgeons' eye gaze} as a holographic marker on the HMDs   
    \item Improve collaboration in OR through providing a multi-user environment with \textbf{interactive patient-specific 3D models of the organ}
    \item Provide \textbf{objective real-time skill assessment} in OR based on task completion time and the endoscopic tool trajectory
     \item Give \textbf{real-time feedback on endoscopic tool manipulation} in the form of holographic instructions to guide the surgical trainee through the procedure when the attending surgeon is not available
     \item Provide \textbf{erasable hologram feature} in OR to allow efficient communication between the trainees and experts by drawing on other holograms or physical components of the surgical setup
     \item Provide video recording of the whole procedure from the expert surgeon perspective using HMDs for \textbf{observational learning} and \textbf{individual or peer-to-peer practice} by surgical trainees
     \item Design visual markers with \textbf{contrastive colors} to the body tissue to visualize expert gaze
     \item Enable \textbf{adjustable size and transparency} control that can allow individuals to adjust their marker size, and hide/unhide the marker from their HMD
\end{itemize}

\subsection{Limitations and Future Work}
A key limitation of our study is that participants were primarily residents from the urology department, with six medical students from urology and otolaryngology. The co-designs generated by residents reflected their expertise, focusing on common OR surgeries, while the medical students’ co-designs centered on pre-clinical training materials. Although residents mainly focused on kidney-related endoscopic surgeries, many of them designed features that can generalize to other surgical specialties. These could provide visual guidance in navigation-based procedures (e.g., bronchoscopy, ureteroscopy, colonoscopy) and surgical education. While these are the ideas generated by our participants, different institutions influence exposure to different training systems, and future research should explore co-designs from surgeons with varied surgical experiences. A study with more trainees could identify AR features useful across disciplines. We emphasize that the features and design considerations solicited from participants are conceptual suggestions, grounded in their domain expertise and lived experience. While these ideas remain theoretical at this stage, such input helps identify user-centered needs, prioritize relevant functionalities, and guide the design of future prototypes. Importantly, these insights offer a foundation for iterative testing and refinement, ensuring that subsequent implementations remain aligned with real-world training challenges. The integration and evaluation of these suggestions will be an essential focus of future work.

Additionally, all 18 participants had used robotic consoles for training, but only 3 had experience with HMDs or VR simulators, possibly limiting the scope of their co-designs. This limited use of AR can lead to a misunderstanding of the capability of the technology, so participant co-designs may not make full use of AR's capabilities. Rather, they demonstrate the needs the trainees express for improvements to surgical training. Future work should incorporate HMDs with AR features for testing in late-stage decision-making activities. Some co-design groups were composed of a single participant, limiting collaboration and potentially introducing a confounding factor. Finally, expert surgeons were not included in the co-design session, and this study does not consider the challenges of training from their perspective.

The primary goal of this work was to generate design implications for improving AR-based systems, so the summative evaluation of system effectiveness was beyond the scope of this study. Future research will focus on implementing the design ideas suggested by our participants to create an initial prototype for an eye-gaze-based visual guidance AR application. We plan to conduct a summative evaluation to assess the usability and effectiveness of this feature in endoscopic procedures.

\section{Conclusion}
In this study, we conducted semi-structured interviews and co-design sessions with surgical trainees to identify gaps in current training methods and explore the integration of AR technology into intraoperative endoscopic training. Participants explored the theoretical integration of AR technology and its potential role in the OR. The participants' solicitations emphasized the value of AR-based assistive tools in the OR and highlighted key design considerations for visual guidance. We developed design considerations for AR features to enhance the collaborative training between surgical experts and trainees. Our work lays the foundation for future studies on AR-based assistive systems in surgical training.

\bibliographystyle{elsarticle-num}
\bibliography{references}

\end{document}